\documentclass[aps,preprint,pra]{revtex4-1}
\usepackage{graphicx}
\usepackage{amssymb}
\usepackage{amsmath}
\usepackage{color}
\usepackage{enumerate}
\usepackage{bm}
\begin{document}
%\iffalse
\title{ Hysteresis in Two Dimensional Arrays of Magnetic Nanoparticles}
\author{Manish Anand}
\email{itsanand121@gmail.com}
\affiliation{Department of Physics, Bihar National College, Patna University, Patna-800004, India.}

\date{\today}

\begin{abstract}
We perform computer simulations to probe the magnetic hysteresis in a two-dimensional ($L^{}_x\times L^{}_y$) assembly of magnetic nanoparticles as a function of dipolar interaction strength $h^{}_d$, temperature $T$, aspect ratio $A^{}_r=L^{}_y/L^{}_x$, and the applied alternating magnetic field's direction. In the absence of magnetic interaction ($h^{}_d\approx0$) and thermal fluctuations ($T=0$ K), the hysteresis follows the Stoner and Wohlfarth model, as expected. For weak dipolar interaction and substantial temperature, the hysteresis has the dominance of superparamagnetic behaviour, irrespective of the applied magnetic field's direction and $A^{}_r$. Interestingly, the hysteresis curve has all the characteristics of antiferromagnetic interaction dominance for $A^{}_r\leq6$ and considerable dipolar interaction strength ($h^{}_d>0.2)$, which is independent of applied magnetic direction. When the magnetic field is applied along the system's shorter axis ($x$-direction), a non-hysteresis straight line is observed with large $h^{}_d$. In the case of the magnetic field applied along the long axis of the sample ($y$-direction), ferromagnetic interaction dominates the hysteresis for large $h^{}_d$ and $A^{}_r>6$. Irrespective of $h^{}_d$ and applied magnetic field's direction, the coercive field $\mu^{}_oH^{}_c$ and remanence $M^{}_r$ are minimal for $A^{}_r\leq6$ and significant temperature. They are found to increase with $h^{}_d$ when $A^{}_r$ is enormous.
%There is a strong dependence of $\mu^{}_oH^{}_c$ and $M^{}_r$ on temperature for weakly interacting MNPs. While for large $h^{}_d$, they depend weakly on temperature. 
Remarkably, the variation of hysteresis loop area $E^{}_H$ as a function of these parameters is the same as that of the coercive field variation. We believe that the concepts presented in this work are relevant in various technological applications such as spintronics and magnetic hyperthermia, in which such self-assembled nanoparticle arrays are ubiquitous.
\end{abstract}
\maketitle
%\fi
\section{Introduction}
In recent years, two-dimensional arrays of magnetic nanoparticles (MNPs) have received significant attention due to their unique magnetic properties and potential technological applications~\cite{gallina2020,alkadour2019,bupathy2019,wang2019,leo2018,wang2008,takagaki2008,kechrakos2002}. From theoretical perspectives, they pose an exciting research topic in which competing interactions, reduced dimensionality, shape anisotropy, etc., lead to novel collective behaviours~\cite{puntes2004,fraerman2002,farhan2013}. On the other hand, such systems are ideal candidates for diverse technological applications such as spintronics, high-density data storages, magnetic hyperthermia, etc.~\cite{valdes2021,frey2010,sarella2014,karmakar2011,mohammadpour2020,takagaki2005}. %{\color{red} Therefore, they are ideal systems to study the magnetic interactions and switching behaviour.} 
%It is therefore of great interest to understand factors which govern the quantitative and qualitative behaviour of TMR.

%Besides promising technological applications, periodic arrays of magnetic dots are the model systems for studying magnetic interactions and switching behavior
%Therefore, it is important to study the magnetization reversal of nanoparticle arrays.    

%From a fundamental perspective, these systems constitute an extremely challenging research topic, in which reduced dimensionality, competing interactions, and disorder merge, leading to novel collective behaviors.

%From a technological standpoint, magnetic nanostructures are most relevant for a variety of applications, particularly in the field of spintronics, memory devices, and high-density data storage [19,20].
The magnetic properties of non-interacting MNPs are well understood~\cite{usov2010,fu2018}. However, they are found to interact in an assembly due to dipolar interaction~\cite{ovejero2016,zhao2018}. The dipolar interaction is long-ranged and anisotropic~\cite{barrera2021}. Further, it can allow for both ferromagnetic and antiferromagnetic couplings depending on the relative position of MNPs. As a consequence, it has varied effects on systemic properties~\cite{kachkachi2000,anand2021thermal,plumer2010}. In some cases, the dipolar interaction causes spins's frustration, modifies energy barriers, and imparts spin-glass-like properties. When the system size is smaller than the range of dipolar interaction, it can create unconventional spins morphologies. Depending on the lattice structure, there have been observations of magnetic vortices, antiferromagnetic states, checkerboard patterns, etc.~\cite{low1994,edlund2010}. Consequently, they exhibit unusual ground-state and
relaxation characteristics. For example, the dipolar interaction elevates the blocking temperature even with weak interaction strength~\cite{luo1991}. The dipolar interaction also affects the ground state spins configuration depending on the geometry of the system. In the face-centred cubic arrangement of MNPs, Luttinger {\it et al.} have found the ground state to be ferromagnetic while it is antiferromagnetic in a simple cubic lattice~\cite{luttinger1946}. The dipolar interaction promotes antiferromagnetic coupling with the square arrangement of MNPs~\cite{macisaac1996,de1997}. On the contrary, the dipolar interaction promotes ferromagnetic coupling in the assembly of magnetic moments arranged in triangular lattice~\cite{politi2006}.

The dipolar interaction also drastically modifies the magnetic properties in such cases~\cite{anand2016,landi2014role,anand2021dipolar,russier2001,xue2006,figueiredo2007}.  For instance, V. Russier studied the magnetic properties, and orientational structure for spherical assembly of MNPs arranged in square and hexagonal lattices at temperature $T=0$ K~\cite{russier2001}. The dipolar interaction is found to induce in-plane orientation of magnetic moments.  Xue {\it et al.} studied the magnetic hysteresis behaviour in a hexagonal array of MNPs with perpendicular anisotropy~\cite{xue2006}. The shape of the hysteresis  curve changes from  rectangular to a non-hysteresis straight line, increasing dipolar interaction strength. For the triangular arrangement of MNPs, Figueiredo {\it et al.} observed an increase in the blocking temperature and relaxation time with increased dipolar interaction strength~\cite{figueiredo2007}.
%Using Monte Carlo simulation, Figueiredo {\it et al.} studied the magnetic relaxation and thermal properties in a triangular arrangment of dipolar interacting MNPs. They observed an increase in the blocking temperature and relaxation time with an increase in dipolar interaction strength~\cite{figueiredo2007}. These observations can be interpreted due to increase in the energy barrier seen by the nanoparticles due to dipolar interaction. 
The dipolar interaction also affects the hysteresis loop area in an assembly of MNPs, one of the most valuable parameters for magnetic hyperthermia applications. Using the molecular dynamics anisotropic diffusion model, Myrovali {\it et al.} have found the dipolar interaction increases the hysteresis loop area in an assembly of interacting MNPs~\cite{myrovali2016}. Using kinetic Monte Carlo simulation, we have shown that dipolar interaction increases the hysteresis loop area in a one-dimensional chain of MNPs~\cite{anand2020}. Using micromagnetic simulations, Haase and Nowak reported that magnetic hyperthermia  efficiency decreases with an increase in dipolar interaction strength~\cite{haase2012}. The dipolar interaction also has a drastic effect on the tunnel magnetoresistance, a useful quantifier for spintronics based applications~\cite{black2000,tan2009,tan2010,kechrakos2008,yang2006}. For example, Tan {\it et al.} found the detrimental effect of dipolar interaction on the tunnel magnetoresistance in square arrays of MNPs~\cite{tan2010}. Kechrakos {\it et al.} and Yang {\it et al.} observed an enhancement in the tunnel magnetoresistance because of dipolar interaction~\cite{kechrakos2008,yang2006}.
%Inoue et al. considered non-interacting MNPs arrays and showed that conductivity is dependent on the relative orientation of magnetic moments of all MNPs~\cite{inoue1996}.}

The above studies clearly suggest that dipolar interaction plays an essential role in determining the magnetic properties of self-assembled MNPs, which are of profound technological importance. 
%Two-dimensional assembly of MNPs are ideal systems to probe the role of dipolar interaction, system size and other parameters of interest on magnetic properties which may play an important role 
%Two dimensional assembly of MNPs has various interesting properties due to geometry confinement, structural order and well defined inparticle magnetic interaction~\cite{}. 
%It is an ideal system to investigate the role of dipolar interactions as undesired complications introduced by spatial disorders are subtantially supressed. {\color{red} motivation, write about probable applications in hyperthermia, spintronics}
However, a systematic study of the effect of dipolar interaction strength, temperature, the shape of the system, and external magnetic field's direction on the underlying system's magnetic properties have received little attention in the literature. These studies could provide a concrete theoretical basis in choosing suitable system parameters for various technological applications of interest. In the present work, we perform kinetic Monte Carlo (kMC) simulations to study the magnetic hysteresis in the two-dimensional assembly of MNPs as a function of various parameters of interest. In particular, we perform a detailed study of the dependence of magnetic hysteresis on dipolar interaction strength, temperature, aspect ratio and direction of the applied magnetic field. 

This paper is organized as follows. In Sec. II, we present the
model for an assembly of MNPs and discuss the kMC simulations briefly. The numerical results will be presented in Sec. III. Finally, we summarize and
discuss the main results in Sec. IV.

\section{Theoretical Framework}
We consider an assembly of spherical magnetic nanoparticles arranged in a two-dimensional lattice in the $xy$-plane with system dimension $L^{}_x\times L^{}_y$. The particle has a diameter $D$, and the lattice spacing is $a$, as shown in Fig.~\ref{figure1}(a). Each particle has a magnetic moment $\mu=M^{}_sV$, $M^{}_s$ being the saturation magnetization , and $V=\pi D^3/6$ is the volume of the nanoparticle. Further, each nanoparticle is assumed to have uniaxial anisotropy $\vec{K}= K^{}_{\mathrm {eff}} \hat{k}$, where $K^{}_{\mathrm {eff}}$ is the uniaxial anisotropy constant, and $\hat{k}$ is the direction of anisotropy or easy axis. We have assumed anisotropy axes to be randomly oriented. The energy associated with such nanoparticle due to magnetocrystalline anisotropy is given by~\cite{carrey2011,anand2018}
\begin{equation}
E^{}_K=K^{}_{\mathrm{eff}}V\sin^2\theta
\end{equation}
Here $\theta$ is the angle between anisotropy axis and magnetic moment.
%\subsection{Aggregation and cluster formation in suspensions of magnetic nanoparticles}
%\label{agg}

MNPs also interact due to the long-ranged dipolar interaction~\cite{pei2019}. %and they form small clusters~\cite{pei2019,chantrell1982}.
%Clustering is relatively common in suspensions of MNPs or magnetic fluids, as observed in electron microscopy or light scattering experiments~\cite{pei2019,chantrell1982}. 
%Normal magnetic fluids contain $\sim10^{23}$ particles/cc and the collisions between particles are frequent. 
%As the particles are magnetised, they adhere and form agglomerates~\cite{pei2019,chantrell1982}. 
We can calculate the dipolar interaction energy $E^{}_{\mathrm {dip}}$ in an assembly of MNPs as~\cite{odenbach2002,rosensweig1997,usov2017}
%\begin{equation}
%\label{dipole}
%E_{dd}(s)= -\frac{1}{4\pi \mu_{o}}\left(\frac{3\left(\vec{\mu_{i}}\cdot\vec{s}\right)\left(\vec{\mu_{j}}\cdot \vec{s}\right)}{s^{5}} -\frac{\vec{\mu_{i}}\cdot \vec{\mu_{j}} }{s^3}\right),
%\end{equation}
\begin{equation}
\label{dipole}
E^{}_{\mathrm {dip}}=\frac{\mu^{}_o}{4\pi a^3}\sum_{j,\ j\neq i}\left[ \frac{\vec{\mu_{i}}\cdot\vec{\mu_{j}}}{(r^{}_{ij}/a)^3}-\frac{3\left(\vec{\mu_{i}}\cdot\hat{r}_{ij}\right)\left(\vec{\mu_{j}}\cdot\hat{r}_{ij}\right)}{(r^{}_{ij}/a)^3}\right].
\end{equation}
Here $\mu_{o}$ is the permeability of free space; $\vec{\mu}_{i}$ and $\vec{\mu}_{j}$ are the magnetic moment vectors of $i^{th}$ and $j^{th}$ nanoparticle respectively, and $r^{}_{ij}$ is the center-to-center separation between $\mu_{i}$ and $\mu_{j}$.  $\hat{r}^{}_{ij}$ is the unit vector corresponding to $\vec{r}_{ij}$.
%The particle has a magnetic moment $\mu=M^{}_sV$, $M^{}_s$ is the saturation magnetization. 
%$\mu^{}_o\vec{H}^{}_{\mathrm {dip}}$ is the dipolar field due to all other nanoparticles, present in the system. 
The corresponding dipolar field $\mu^{}_o\vec{H}^{}_{\mathrm {dip}}$ is given by the following expression~\cite{rosensweig1997,tan2014}
\begin{equation}
\mu^{}_{o}\vec{H}^{}_{\mathrm {dip}}=\frac{\mu\mu_{o}}{4\pi a^3}\sum_{j,j\neq i}\frac{3(\hat{\mu}^{}_j \cdot \hat{r}_{ij})\hat{r}^{}_{ij}-\hat{\mu^{}_j} }{(r_{ij}/a)^3}.
\label{dipolar1}
\end{equation}
%Here $\mu_{j}$ is the magnetic moment vector of the $j^{th}$ particle, $s$ is the center-to-center separation between $\mu_{i}$ and $\mu_{j}$; and the $\mu_{o}$ is the permeability of free space.
As it is clearly evident from Eq.~(\ref{dipole}) and Eq.~(\ref{dipolar1}) that dipolar interaction varies as $1/r^{3}_{ij}$, we can define the strength of dipolar interaction $h^{}_d=D^{3}/a^3$~\cite{tan2010}. So $h^{}_d= 1.0$ is the largest dipolar interaction
strength, and $h^{}_d=0$ can be termed as the non-interacting case.

We apply an oscillating magnetic field to probe the hysteresis behaviour of dipolar interacting MNPs. It is given by~\cite{anand2020}
%To study the magnetic hysteresis, we apply an alternating magnetic field ${\mu^{}_{o}H}$ along the $z$-direction (along the chain axis of MNPs) expressed as
\begin{equation}
\mu^{}_{o}H=\mu^{}_oH^{}_{\mathrm {o}}\cos\omega t,
\label{magnetic}
\end{equation} 
where $\mu^{}_{o}H_{\mathrm {o}}$ and $\omega=2\pi\nu$ are the amplitude and angular frequency of the applied magnetic field respectively, $\nu$ is the linear frequency, and $t$ is the time. Therefore, the total energy of the underlying system under the influence of dipolar and external magnetic field is given by~\cite{tan2014,anand2019}
\begin{equation}
E=K^{}_{\mathrm {eff}}V\sum_{i}\sin^2 \theta^{}_i+\frac{\mu^{}_o\mu^2}{4\pi a^3}\sum_{j,\ j\neq i}\left[ \frac{\hat{\mu_{i}}\cdot\hat{\mu_{j}}-{3\left(\hat{\mu_{i}}\cdot\hat{r}_{ij}\right)\left(\hat{\mu_{j}}\cdot\hat{r}_{ij}\right)}}{(r_{ij}/a)^3}\right]-\mu^{}_o\sum_{i}\vec{\mu}^{}_i\cdot\vec{H}
\end{equation}
Here $\theta^{}_i$ is the angle between the anisotropy axis and the $i^{th}$ magnetic moment of the system. 
%and $\phi^{}_i$ is the angle between the anisotropy field and the total magnetic field $ H^{}_{\mathrm {total}}$ (dipolar and external field).

%It is a well known fact that MNPs organize in linear chain  geometry under the influence of magnetic field~\cite{chantrell1982,morimoto2009,usov2020}. To ascertain it, we first identify low energy geometric configurations of the tiny cluster (in the absence of an oscillating magnetic field) using first-principle calculation.

We implement kMC simulations to probe the hysteresis response as a function of dipolar interaction strength $h^{}_d$, temperature $T$, aspect ratio $A^{}_r=L_y/L_x$, and the applied magnetic field's direction. In the kMC algorithm, dynamics is captured more accurately, which is essential to study the dynamic hysteresis response of dipolar interacting MNPs~\cite{chantrell2000}.
%study the magnetic hysteresis of dipolar interacting MNPs in a linear chain as a function of the orientation of the anisotropy axis, dipolar interaction strength, frequency, and temperature.  
%In this algorithm, dynamics is more accurately taken into account, which is necessary to study the magnetic hysteresis properties at high frequency. {\color{blue} The upper limit of frequency that the kMC simulations can be used is when the frequency is smaller than the precession of the magnetization time scale,   usually $\sim 10^{-9}$ s.}
Using this technique, we can accurately describe the dynamical properties of MNPs in the superparamagnetic or ferromagnetic regime without any artificial or abrupt separation between them~\cite{tan2014}. We have used the same algorithm, which is described in greater detail in the work of Tan {\it et al.} and Anand {\it et al.}~\cite{tan2014,anand2019}. Therefore, we do not reiterate it here to avoid repetition.

The amount of heat dissipated by the MNPs assembly during one complete cycle of the external magnetic field equals the hysteresis loop area $E^{}_{H}$. It can be evaluated using the following formula~\cite{anand2016}
\begin{equation}
E^{}_{H}=\oint M(H)dH,
\label{local_heat}
\end{equation}
The above intergral is evaluated over the entire period of the external magnetic field change. $M^{}(H)$ is the magnetization of system at magnetic field $H$.

\section{Simulations Results}
We consider spherically shaped nanoparticles of Fe$_3$O$_4$ arranged in a two-dimensional lattice ($L^{}_x\times L^{}_y$) with $D=8$ nm, $K_{\mathrm {eff}}=13\times10^3$ Jm$^{-3}$, and $M^{}_s=4.77\times10^5$ Am$^{-1}$. The total number of nanoparticles in the assembly is considered as $n=400$. We have considered seven values of system sizes viz. $L_x\times L_y=20\times20$, $16\times25$, $10\times40$, $8\times50$, $4\times100$, $2\times200$ and $1\times400$. The corresponding aspect ratio $A^{}_r(=L^{}_y/L^{}_x)$ of the underlying system is $1.0$, 1.56, 4.0, 6.25, 25, 100 and 400, respectively. The dipolar interaction strength $h^{}_d$ is varied from 0 to 1.0. The temperature $T$ is changed between 0 and 300 K. The anisotropy axes are assumed to have random orientations. The alternating magnetic field is applied along $x$ and $y$-direction. The amplitude of the magnetic field strength $\mu^{}_oH^{}_{\mathrm {o}}$ and its frequency $\nu$ is taken as $0.10$ T and $10^5$ Hz, respectively. 
%It means that the system is in the non-linear reponse regime as $\mu_oH_o>H^{}_K$, $H^{}_K=2K_{\mathrm {eff}}/M^{}_s$ is single particle anisotropy field.
%The frequency of the applied field $\nu$ is taken as $10^5$ Hz. The amplitude of the field strength is taken as $\mu^{}_oH^{}_{\mathrm {max}}= 0.10$ T. 

\begin{figure}[!htb]
\centering\includegraphics[scale=0.60]{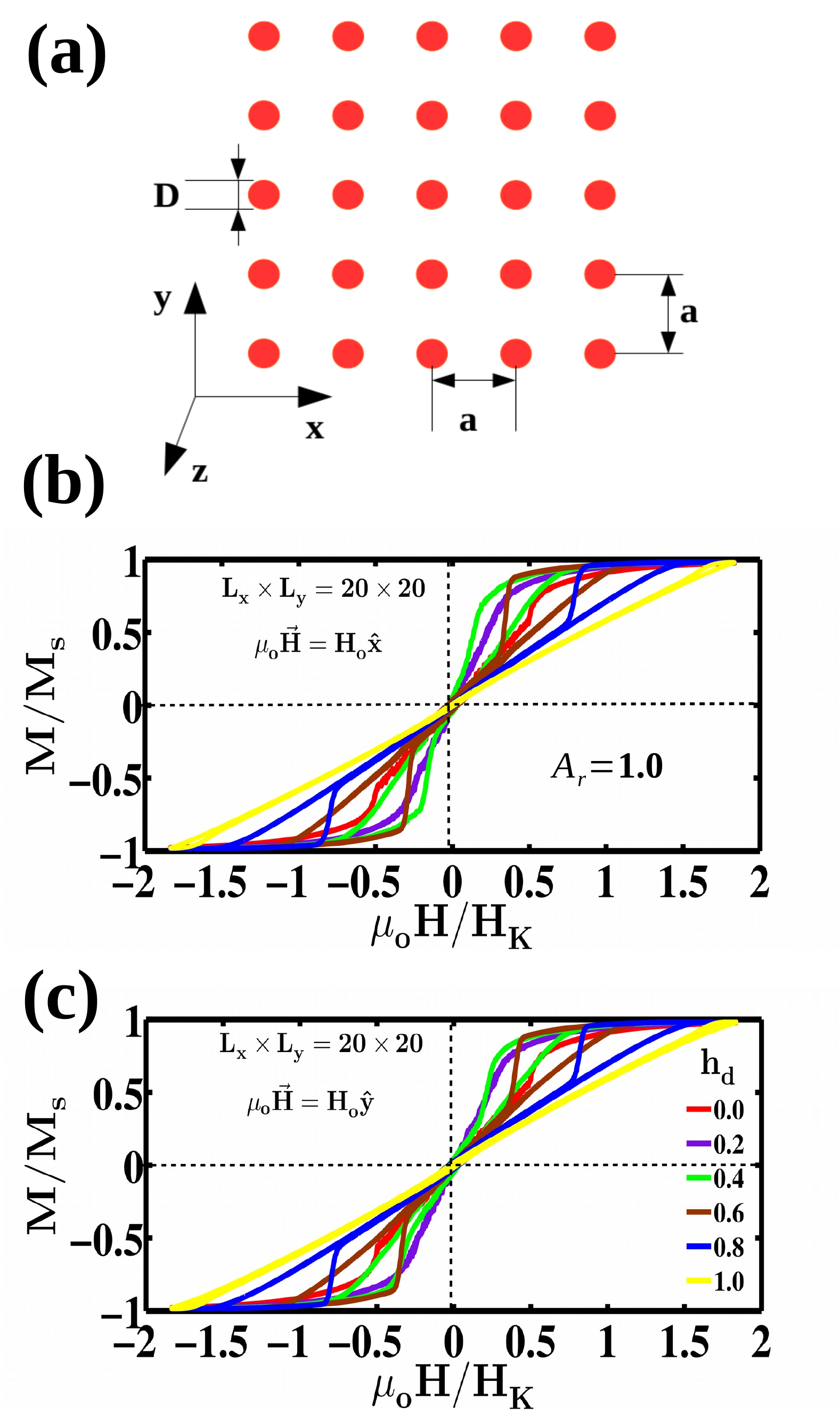}
\caption{(a) Schematic of the two-dimensional assembly of MNPs used in the present work, $D$ is the particle diameter and $a$ the lattice spacing. (b) Magnetic hysteresis curves for MNPs arranged in a square lattice as a function of dipolar interaction strength at $T=300$ K. The external alternating magnetic field is applied along $x$-direction [(b)] and $y$-direction [(c)] at $T=300$ K. For non-interacting and weakly interacting MNPs ($h^{}_d\leq0.2$), the hysteresis loop area is minimal, indicating the dominance of superparamagnetic behaviour. Interestingly, the hysteresis curve has all the antiferromagnetic interaction characteristics for considerable dipolar interaction strength ($h_d>0.2$). The hysteresis response is independent of
the applied magnetic field's direction (in-plane) in a square lattice.}
\label{figure1}
\end{figure}

First, we study the magnetic hysteresis in a square lattice with $L^{}_x\times L_y=20\times20$. we have considered six representative values of dipolar interaction strength $h^{}_d=0$, 0.2, 0.4, 0.6, 0.8 and 1.0. The temperature is taken as $300$ K, and the magnetic field is applied along $x$-direction in Fig.~\ref{figure1}(b) and along $y$-direction in Fig.~\ref{figure1}(c). In all the hysteresis curves shown in the present article, the $x$-axis (applied magnetic field) and $y$-axis (magnetization) have been scaled by single-particle anisotropy field $H^{}_K=2K_{\mathrm {eff}}/M^{}_s$~\cite{carrey2011} and $M^{}_s$, respectively. The hysteresis curve has the negligible value of coercive field and remanence for weakly interacting MNPs ($h^{}_d\leq 0.2$) indicating the dominance of superparamagnetic character. Interestingly, the dipolar interaction induces antiferromagnetic coupling for large magnetic interaction strength ($h_d>0.2$). As a consequence, the corresponding hysteresis curve has all the signatures of the dominance of antiferromagnetic interaction, i.e., zero value of remanence. These observations are in perfect agreement with the work of Chen {\it et al.}~\cite{chen2017}. It is also in qualitative agreement with the work of Ewerlin {\it et al.}~\cite{ewerlin2013}. These observations are also independent of the direction of the applied magnetic field as along it is applied in a plane of the system.

%the hysteresis has the charactersitics of antiferromagnetic interaction for large dipolar interaction strength ($0.2<h^{}_d<0.9$). 

%While for largest interaction strength $h^{}_d$, non-hystersis straight line is observed. 

%Similar observation can be made when the field is applied along $y$-direction [see Fig.~\ref{figure1}(c)].

\begin{figure}[!htb]
\centering\includegraphics[scale=0.50]{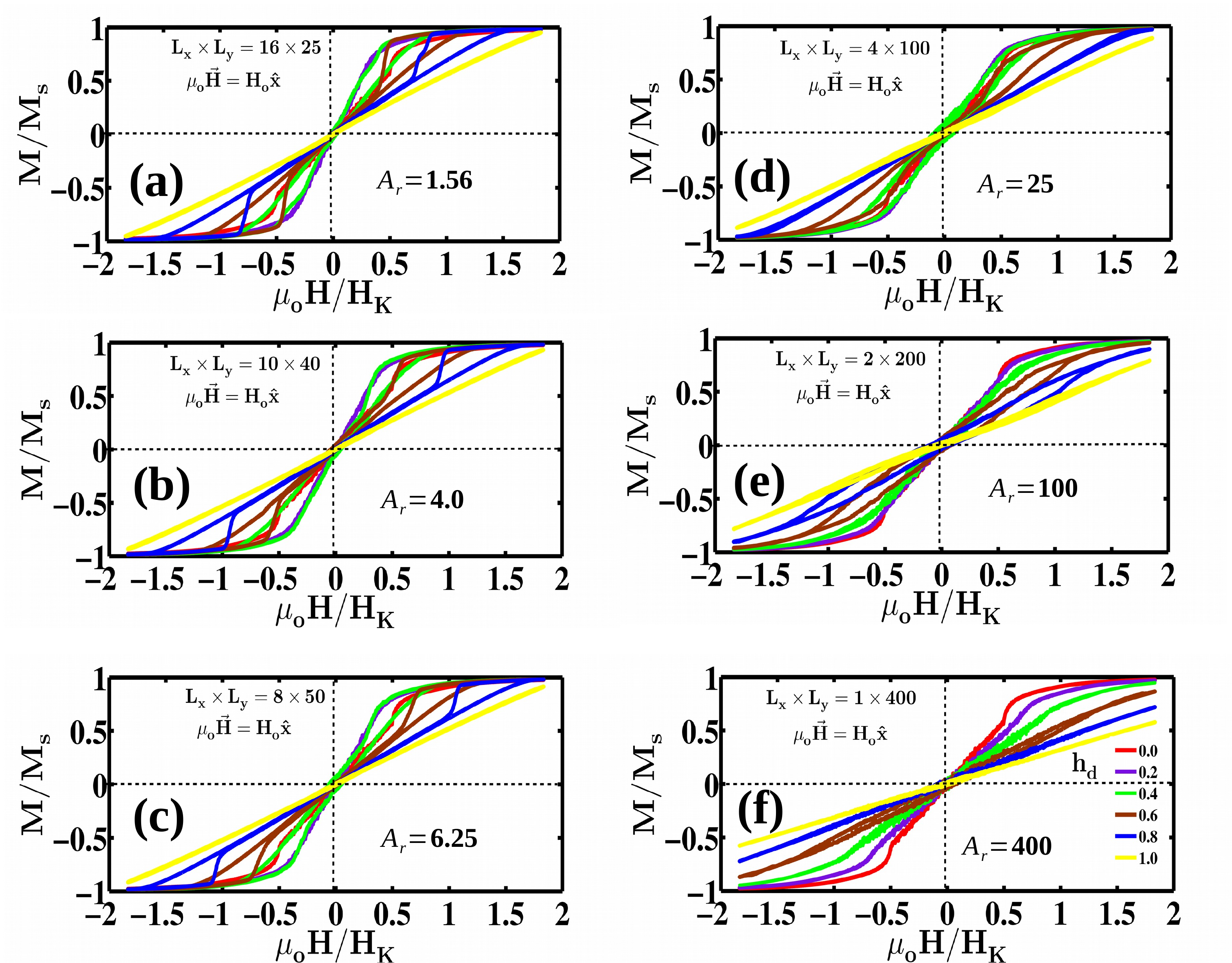}
\caption{Magnetic hysteresis curves as a function of dipolar interaction for six values of aspect ratio $A^{}_r= 1.56$ [(a)], 4.0 [(b)], 6.25 [(c)], 25.0 [(d)], 100 [(e)], and 400 [(f)] at $T=300$ K. The dipolar interaction strength $h^{}_d$ is varied from 0 to 1.0 and the alternating magnetic field is applied along the $x$-direction. There is a negligible hysteresis loop area with small $h^{}_d$, which indicates the dominance of superparamagnetic behaviour. For large dipolar interaction strength and $A^{}_r\leq6.25$, the hysteresis curve has a negligible value of remanence and coercive field, which indicates the dominance of antiferromagnetic coupling. Non-hysteresis is observed for extremely large aspect ratio ($A^{}_r>6.25$) and dipolar interaction strength $h_d>0.8$.}
\label{figure2}
\end{figure}

To see the effect of aspect ratio $A^{}_r$ and direction of the applied magnetic field on the hysteresis, we then study magnetic hysteresis for six values of $A^{}_r$ in Fig.~(\ref{figure2}). The magnetic field is applied along the $x$-direction, and temperature $T$ is taken as 300 K. Irrespective of $A^{}_r$, the hysteresis loop area is negligibly tiny for very weak dipolar interaction strength ($h^{}_d\leq0.2$), indicating the dominance of superparamagnetic behaviour. For large dipolar interaction strength and relatively small aspect ratio $A^{}_r\leq 6.0$, the dipolar interaction induces antiferromagnetic coupling, which is reflected in the negligible value of remanence and coercive field. When the aspect ratio is large ($A^{}_r>6.0$), we observe non-hysteresis for considerable dipolar interaction strength. We can explain it using the fact that the magnetic moment's natural tendency is to get aligned along the sample's longer axis
(along $y$). The applied magnetic field forces the magnetic moment to aligned along the shorter side of the sample. As a consequence, non-hysteresis is observed for sizeable dipolar interaction strength.

\begin{figure}[!htb]
\centering\includegraphics[scale=0.50]{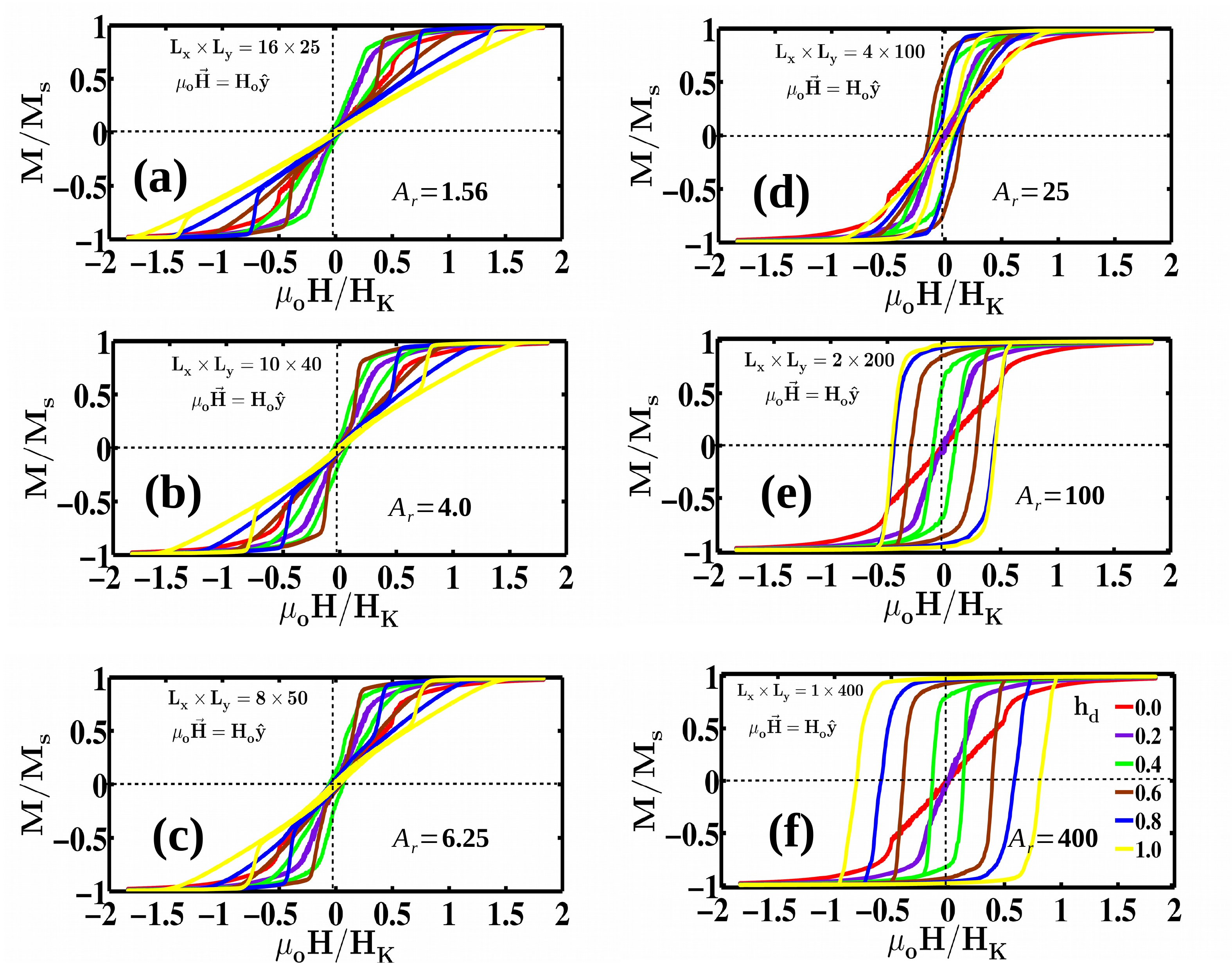}
\caption{Magnetic hysteresis curves as a function of dipolar interaction strength $h^{}_d$ with the magnetic field is applied along the $y$-direction, and $T=300$ K. We have considered six values of aspect ratio $A^{}_r= 1.56$ [(a)], 4.0 [(b)], 6.25 [(c)], 25.0 [(d)], 100 [(e)], and 400 [(f)]. For $A^{}_r\leq6.25$ and large dipolar interaction strength $h^{}_d$, antiferromagnetic interaction dominates the magnetic hysteresis, i.e., the negligible remanence value. Interestingly for large aspect ratio and appreciable dipolar interaction strength, the hysteresis curve has a very large value of coercive field, and remanence. They get enhanced with an increase in $h^{}_d$. These observations indicate the dominance of ferromagnetic interaction.}
\label{figure3}
\end{figure}

In Fig.~(\ref{figure3}), we apply the magnetic field along the $y$-direction. All other parameters are the same as that of Fig.~(\ref{figure2}). There is a dominance of antiferromagnetic interaction for considerable dipolar interaction strength ($h_d>0.2$) and $A^{}_r<6.0$ in this case also. While for weak dipolar interaction strength, the hysteresis loop area is negligibly small irrespective of $A^{}_r$, which indicates the signature of superparamagnetic behaviour. Remarkably, the dipolar interaction induces ferromagnetic coupling for $h_d>0.2$ and a very large  aspect ratio ($A^{}_r>6.0$). Consequently, the coercive field and remanence have huge values and get enhanced with an increase in $h^{}_d$. One can explain it using the fact that the dipolar interaction forces the magnetic moment to get aligned along the sample's longer axis due to the increased value of shape anisotropy. For extremely large $A^{}_r$ and strongest interacting MNPs ($h^{}_d=1.0$), remanence is close to 1.0, indicating extremely large ferromagnetic coupling. It is evident that for sizeable dipolar interaction strength, there is a transition point at $A^{}_r\leq 6.0$; below this value, antiferromagnetic coupling dominates. For a larger value of $A^{}_r$, ferromagnetic coupling prevails. These results suggest that the nature of the interaction can be tuned from antiferromagnetic to ferromagnetic just by changing the aspect ratio provided the dipolar interaction strength is very large. It has got profound applications in spintronics based applications where one needs large tunnel magnetic resistance, which could be maximized with
the antiferromagnetic arrangement of MNPs~\cite{bupathy2019}. The latter can thus be achieved for the sample having a smaller aspect ratio and sizeable dipolar interaction.

\begin{figure}[!htb]
\centering\includegraphics[scale=0.40]{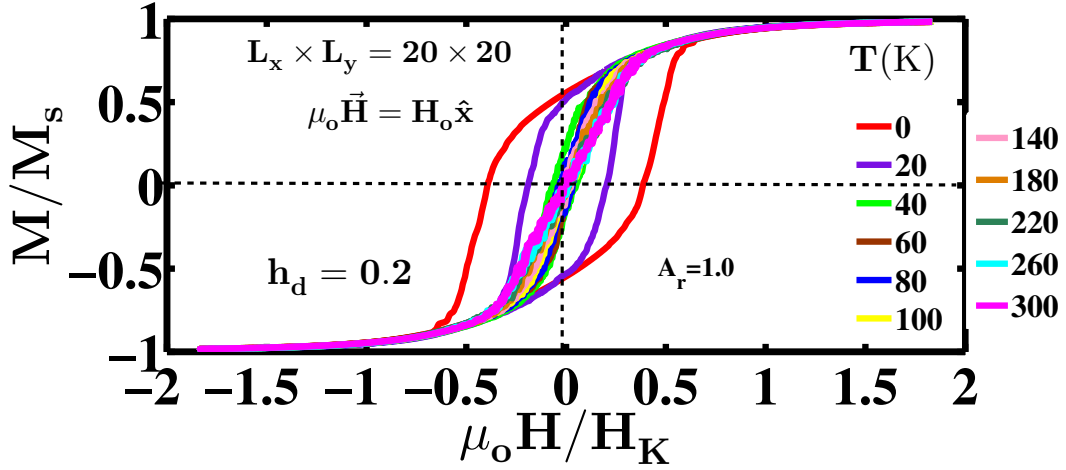}
\caption{Magnetic hysteresis curve as a function of temperature $T$ in  a square arrangement of MNPs ($A^{}_r=1.0$) for weakly dipolar interacting ($h^{}_d=0.2$). The alternating magnetic field is applied along the $x$-direction. In the absence of thermal fluctuations ($T=0$ K), the hysteresis curve has the coercive field $\mu_oH_c\sim 0.48H_K$ and remanence $M^{}_r\sim 0.5$ as expected, which is in perfect agreement with the Stoner and Wohlfarth model. The hysteresis loop area decreases with an increase in temperature because of enhancement in thermal fluctuations. These observations are independent of applied magnetic field's direction (in-plane) and the aspect ratio $A^{}_r$ (curves  not shown).}
\label{figure4}
\end{figure}

Temperature also plays an essential role in determining the magnetic hysteresis response in an array of dipolar interacting MNPs. So, now we study the temperature dependence of hysteresis for various aspect ratio and the applied magnetic field direction. In Fig.~(\ref{figure4}), we plot the hysteresis curve for the square sample ($L^{}_x\times L^{}_y=20\times20$) with various values of $T$ for weakly interacting MNPs ($h^{}_d=0.2$). The magnetic field is applied along the $x$-direction. In the absence of thermal fluctuations ($T=0$ K), the system behaves like a blocked state. The corresponding hysteresis curve shows the coercive field value of $0.48H^{}_K$ and remanence $\sim 0.5$, which are in perfect agreement with the Stoner and Wohlfarth model~\cite{stoner1948}. It is also in qualitative agreement with the work of V. Russier~\cite{russier2001}.
There is a decrease in the hysteresis loop area with an increase in temperature. It can be explained using the fact the with an increase in thermal fluctuations; magnetic moments tends to orient randomly, resulting in the minimal value of coercive field and remanence  provided the dipolar interaction strength is weak. These observations are robust with respect to the aspect ratio and direction of the applied magnetic field. Therefore curves are not shown for other values of $A^{}_r$ and direction of the applied magnetic field to avoid duplications.
 
\begin{figure}[!htb]
\centering\includegraphics[scale=0.50]{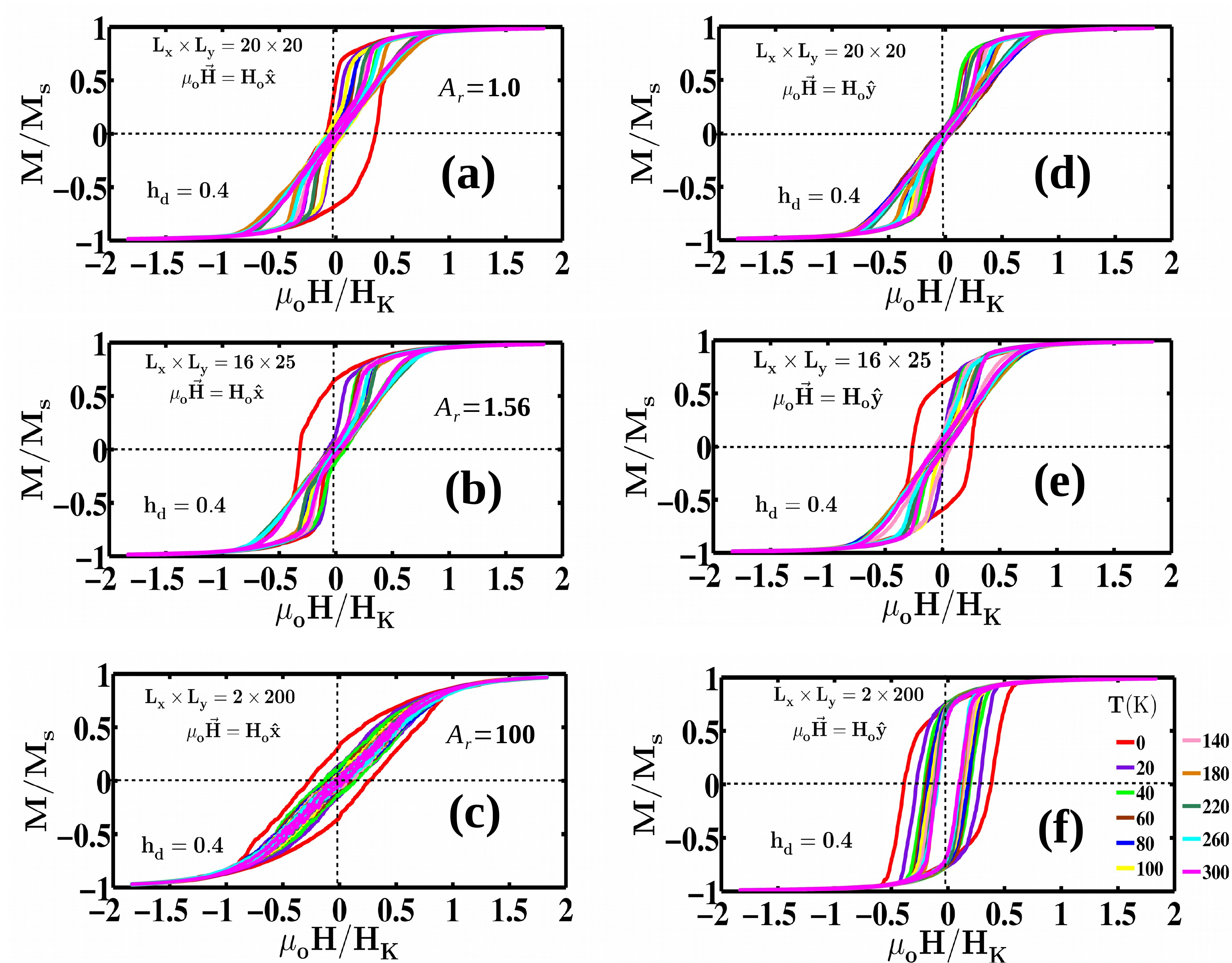}
\caption{Magnetic hysteresis curve as a function of temperature $T$ for moderate dipolar interaction ($h_d=0.4$) and three representative values of aspect ratio $A^{}_r=1.0$, 1.56 and 100. The magnetic field is applied along $x$-direction [(a), (b) and (c)] and $y$-direction [(d), (e) and (f)]. In the case of small $A^{}_r$ and non-zero temperature, antiferromagnetic interaction is dominant resulting in negligible remanence and coercive field, which is independent of direction of the applied magnetic field. While for large $A^{}_r$, the hysteresis curve has small hysteresis loop area with the field applied along the shorter axis of the sample ($x$-direction). Interestingly, there is a large coercive field and remanence with the magnetic field applied along $y$-direction and $A^{}_r$ is very large, which indicates the dominance of ferromagnetic coupling. There is always a decrease in hysteresis loop area with temperature.} 
\label{figure5}
\end{figure}

In the case of sizeable dipolar interaction strength, the hysteresis response should depend on the direction of the applied magnetic field and aspect ratio for a given temperature. Therefore, we study the magnetic hysteresis with $h^{}_d=0.4$ as a function of temperature for three representative values of aspect ratio $A^{}_r=1.0$, 1.56 and 100 in Fig.~(\ref{figure5}). $T$ is varied between 0 to 300 K and the magnetic field along $x$ and $y$-direction. For small $A^{}_r$ and $T>20$ K, antiferromagnetic interaction is dominant resulting in negligible remanence and the coercive field value. The system tends to behave as a blocked state for the small value of $T$. These observations are independent of applied magnetic field's direction. On the other hand, in the case of huge aspect ratio, the hysteresis curve depends strongly on the applied magnetic field's direction. The hysteresis loop area is very small when the field is applied along the $x$-direction. The dipolar interaction induces ferromagnetic coupling when the field is applied along the long axis of the sample ($y$-direction), which results in a tremendous value of coercive field and remanence. Consequently, the hysteresis loop area is very large compared to the field applied along the sample's shorter axis ($x$-direction). The hysteresis loop decreases with an increase in temperature because of enhancement in thermal fluctuations.

\begin{figure}[!htb]
\centering\includegraphics[scale=0.50]{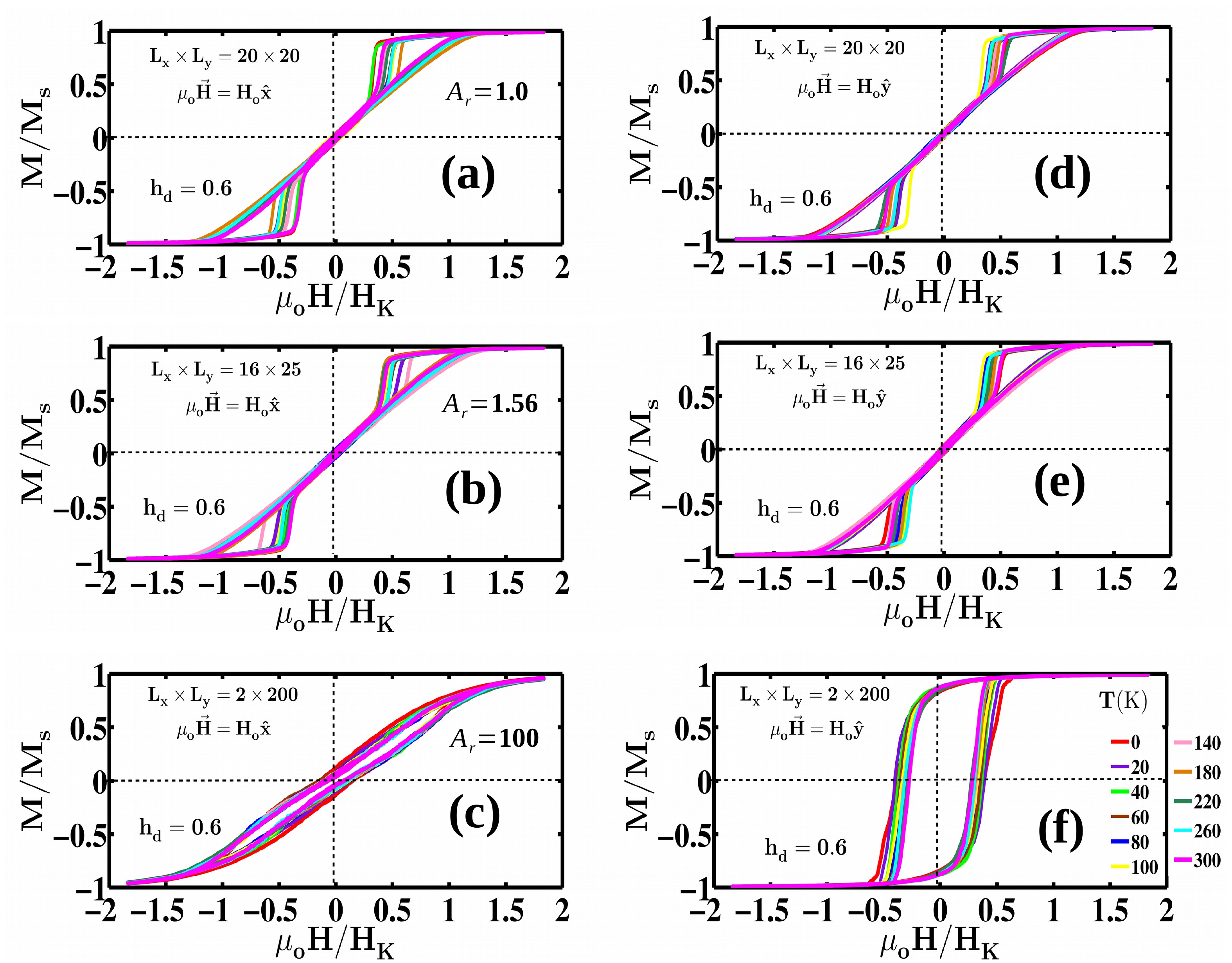}
\caption{Temperature dependence of magnetic hysteresis curve with large dipolar interaction $h_d=0.6$. We have considered three typical values of aspect ratio $A^{}_r=1.0$, 1.56 and 100. The magnetic field is applied is applied along $x$-direction [(a), (b) and (c)] and $y$-direction [(d), (e) and (f)]. Irrespective of the applied magnetic field's direction, the hysteresis curve has all the characteristics of antiferromagnetic interaction for small $A^{}_r$. There is a large hysteresis loop area with the magnetic field applied along the sample's long axis ($y$-direction) and extremely large $A^{}_r$. There is a weak dependence of hysteresis on temperature.}
\label{figure6}
\end{figure}

\begin{figure}[!htb]
\centering\includegraphics[scale=0.50]{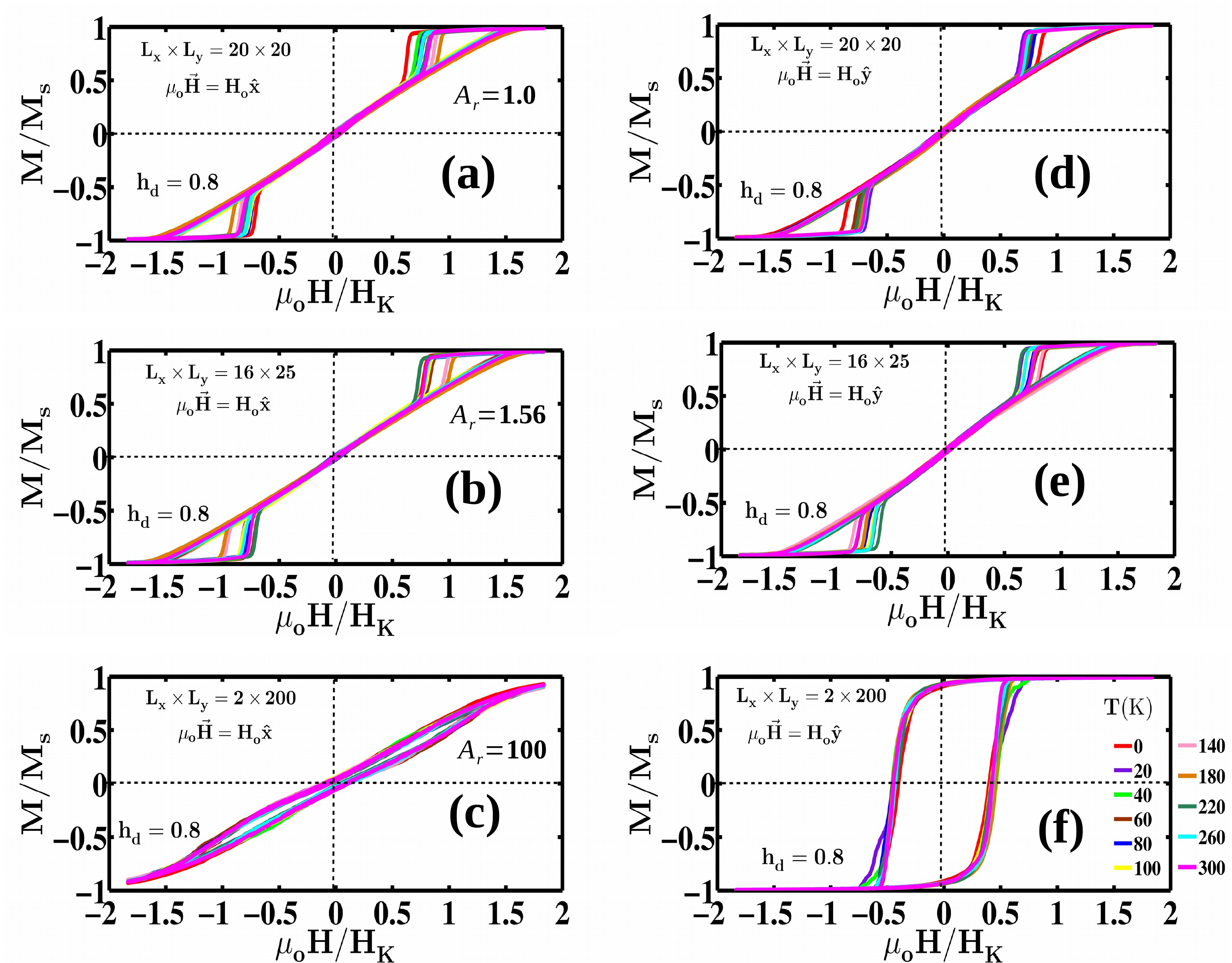}
\caption{Magnetic hysteresis as a function of temperature for very large dipolar interaction strength $h_d=0.8$. We have considered there representative values of $A^{}_r=1.0$, 1.56 and 100; the magnetic field is applied along $x$-direction [(a), (b) and (c)] and $y$-direction [(d), (e) and (f)]. In the case of small $A^{}_r$ and irrespective of applied magnetic field's direction, the hysteresis curve has negligible remanence and the coercive field, indicating the dominance of antiferromagnetic coupling. The hysteresis loop area is minimal with the magnetic field applied along the sample's shorter axis ($x$-direction), provided the aspect ratio is huge. The hysteresis loop area is very large when the direction of the applied magnetic field is along the longer axis of the system ($y$-direction) and $A^{}_r$ is enormous.} 
\label{figure7}
\end{figure}

To see further the effect of temperature on the hysteresis for enormous dipolar interaction strength, we plot magnetic hysteresis curves for $h^{}_d=0.6$ and 0.8 in Fig.~(\ref{figure6}) and Fig.~(\ref{figure7}), respectively. All other parameters are the same as that of Fig.~(\ref{figure5}). Irrespective of the applied magnetic field's direction, the hysteresis curve shows the dominance of antiferromagnetic interaction for a small value of $A^{}_r(\leq6.0)$.
%As dipolar interaction strength has very large value, temprature does not have siginficat impact on the hysteresis behaviour. 
The hysteresis curve has a negligibly small value of the coercive field and remanence when the magnetic field is applied along the sample's shorter axis ($x$-axis). It is due to the fact that the natural tendency of magnetic moments is to get aligned along the long axis of the sample; the applied magnetic field along $x$-direction forces them to orient along the system's shorter axis, which results in a minimal hysteresis loop area. In the case of the magnetic field applied along the long axis of the sample ($y$-axis), the hysteresis loop area is huge, indicating the ferromagnetic coupling dominance. As the dipolar interaction strength is very large, the temperature does not affect the hysteresis behaviour irrespective of the applied magnetic field's direction.

\begin{figure}[!htb]
	\centering\includegraphics[scale=0.50]{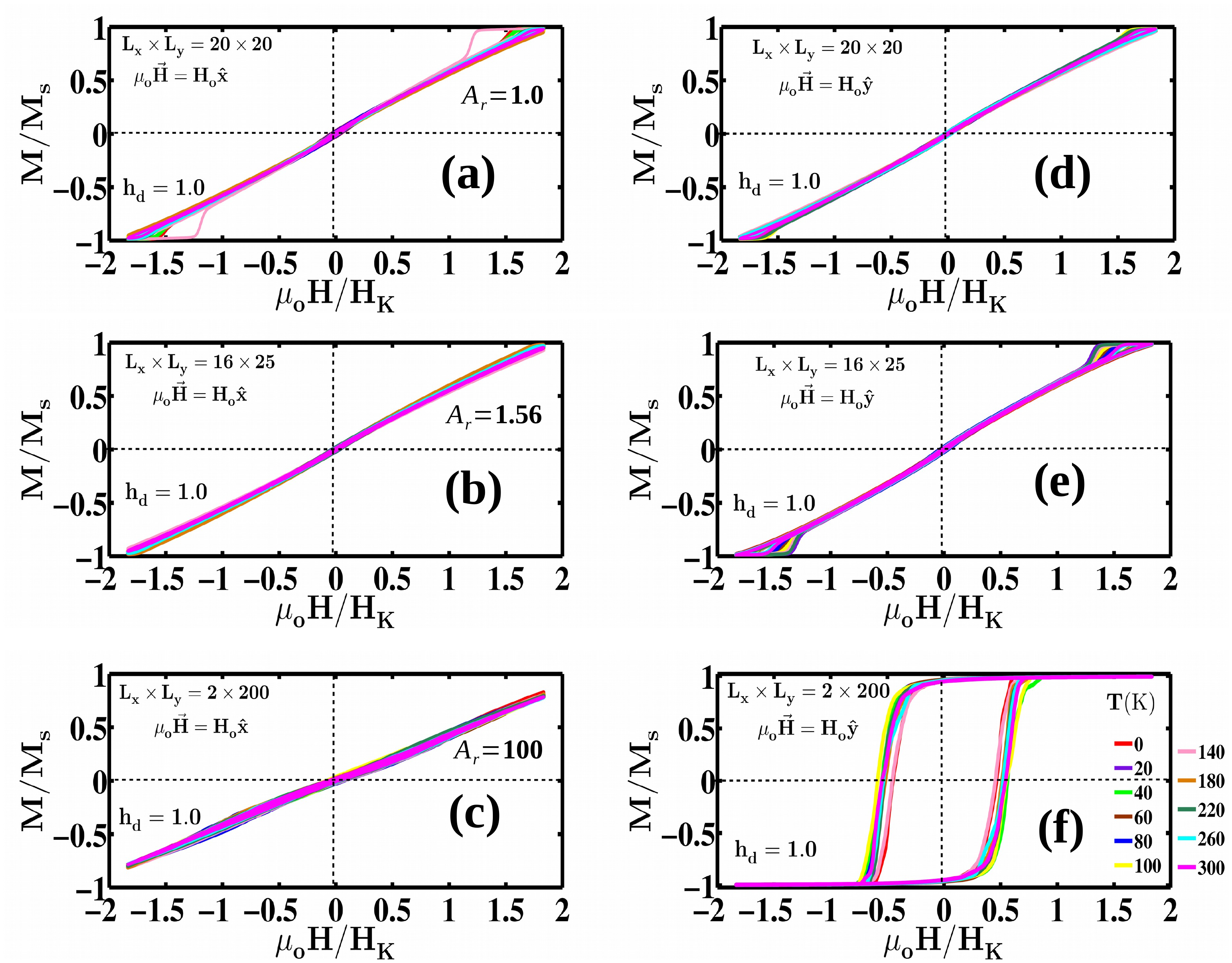}
	\caption{Magnetic hysteresis curves as a function of temperature for the strongest dipolar interacting MNPs ($h^{}_d=1.0$) with three values of $A^{}_r=1.0$, 1.56 and 100. The magnetic field is applied along $x$-direction [(a), (b) and (c)] and $y$-direction [(d), (e) and (f)]. In the case of the magnetic field applied along the $x$-direction, non-hysteresis is observed irrespective of aspect ratio. The hysteresis loop area is huge, indicating the dominance of ferromagnetic interaction when the field is applied to the longer axis of the sample, provided the aspect ratio is enormous. Temperature does not affect the hysteresis because of substantial large dipolar interaction.} 
	\label{figure8}
	\end{figure}

In Fig.~(\ref{figure8}), we study the temperature dependence of hysteresis for the strongest dipolar interacting MNPs ($h^{}_d=1.0$). All other parameters are the same as that of Fig.~(\ref{figure7}). The hysteresis curve has a negligible value of hysteresis loop area irrespective of the applied magnetic field's direction for $A^{}_r\leq6.0$. We observe a non-hysteresis straight line for $A^{}_r$'s very large value with the applied field is along the shorter side of the sample($x$-direction). Remarkably, the hysteresis shows the dominance of ferromagnetic interaction for an extremely large value of $A^{}_r$ when the field is applied along the long axis of the system. In this case, the hysteresis loop area is enormous. The extremely large value of remanence ($\sim 1.0$) also strengthens the fact the ferromagnetic coupling is the maximum. Temperature does not have an impact  on the hysteresis as dipolar interaction strength is the largest in this case.

\begin{figure}[!htb]
	\centering\includegraphics[scale=0.50]{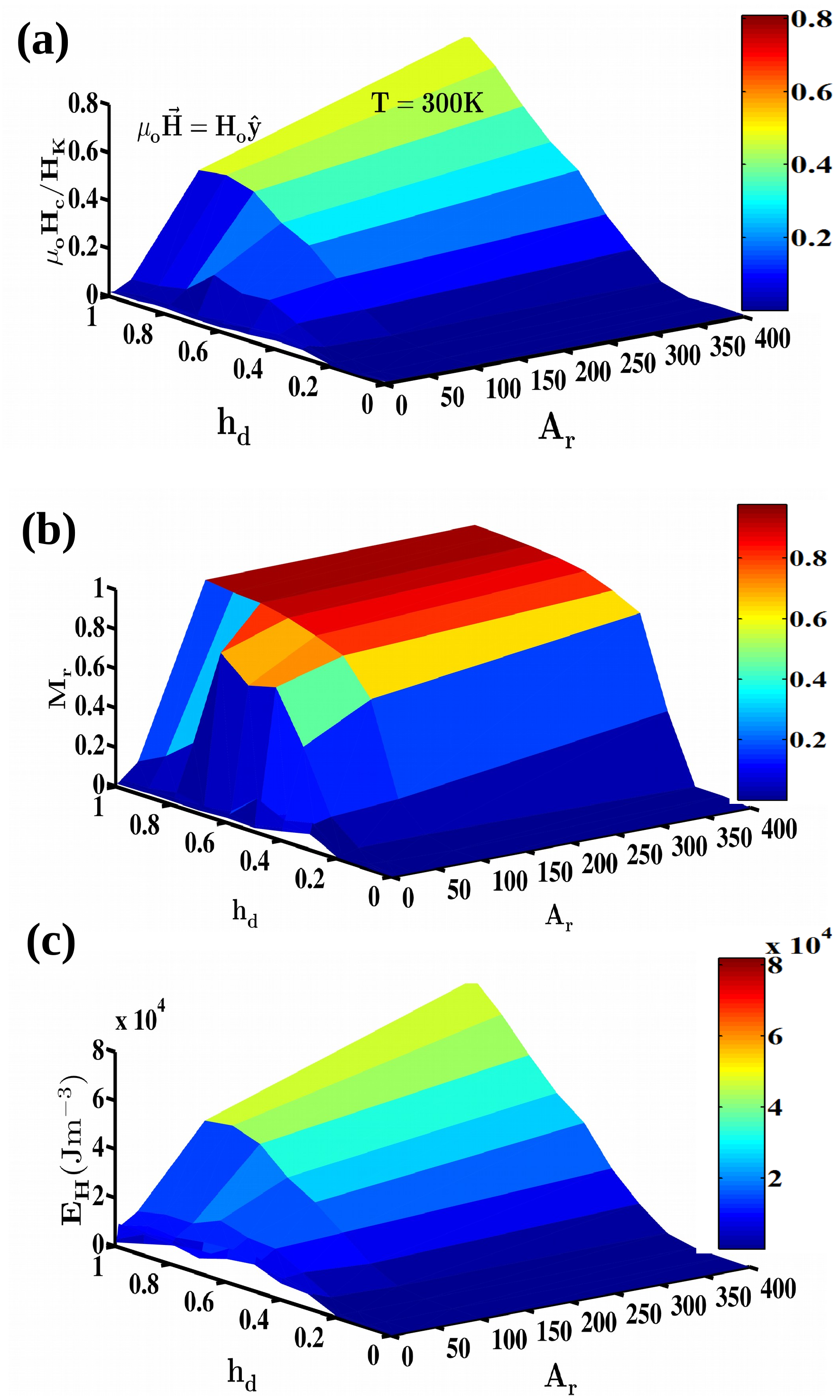}
	\caption{Variation of the coercive field $\mu_oH_c$ (scaled by single-particle anisotropy field $H^{}_K$), remanence $M^{}_r$ and hysteresis loop area $E^{}_H$ as a function of dipolar interaction strength $h^{}_d$ and aspect ratio $A^{}_r$. The magnetic field is applied along the $y$-direction, and temperature $T$ is taken as 300 K. For small aspect ratio and appreciable $h^{}_d$, $\mu^{}_oH^{}_c$ and $M^{}_r$ have negligible values. There is an increase in value of $\mu^{}_oH^{}_c$ and $M^{}_r$ with an increase in $h^{}_d$, indicating ferromagnetic behaviour. Interestingly, the variation of $E^{}_H$ with $h^{}_d$ and $A^{}_r$ is the same as that of coercive field variation.} 
	\label{figure9}
\end{figure}

To quantify the magnetic hysteresis, we now study the variation of coercive field $\mu_oH^{}_c$ (scaled by $H^{}_K$), remanence $M^{}_r$ and hysteresis loop area $E^{}_H$ as a function of aspect ratio $A^{}_r$ and dipolar interaction strength $h^{}_d$ in Fig.~(\ref{figure9}). The magnetic field is applied along the $y$-direction and temperature is taken as $T=300$ K.
The coercive field and remanence are negligibly small for weakly interacting MNPs and small value of $A^{}_r$ indicating the dominance of superparamagnetic behaviour. The antiferromagnetic coupling between the magnetic moment is favoured for large $h^{}_d$ and small values of $A^{}_r$, which is manifested in the minimal value of $M^{}_r$ and $\mu_oH^{}_c$. On the other hand, there is an enhancement in the coercive field and remanence magnetization with an increase in dipolar interaction strength for large $A^{}_r$, indicating the ferromagnetic coupling dominance. When the dipolar interaction strength and aspect ratio are enormous, there is weak dependence of coercive field and remanence on $A^{}_r$. Remarkably, the variation of $E^{}_H$ with aspect ratio and dipolar interaction strength is precisely similar to that of coercive field variation. Our results for $h^{}_d=0$ is in perfect aggreement with the works of Carrey {\it et al.}~\cite{carrey2011}. We could not compare the entire range of aspect ratio and $h^{}_d$ as they have concentrated only on the non-interacting case ($h^{}_d=0$). Therefore, our results can be used as a benchmark in this context. These results could be used for choosing correct values of dipolar interaction strength, aspect ratio for optimizing heat dissipation and tunnel magnetoresistance for magnetic hyperthermia and spintronics based applications.

\begin{figure}[!htb]
	\centering\includegraphics[scale=0.45]{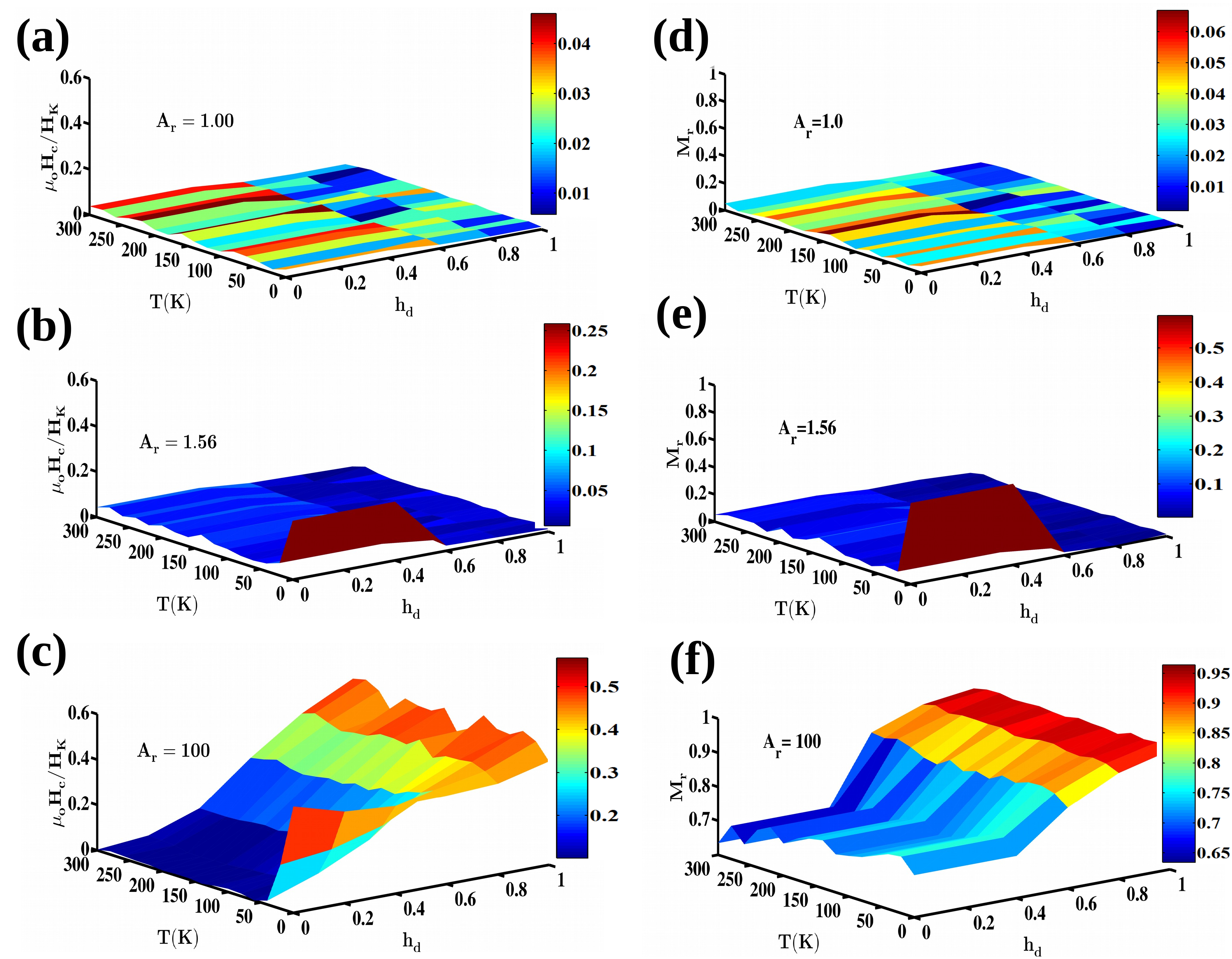}
	\caption{Variation of the coercive field $\mu^{}_oH^{}_c$ (scaled by $H^{}_K$) and remanence $M^{}_r$ as a function of temperature $T$ and dipolar interaction strength $h^{}_d$ for three representative values of aspect ratio $A^{}_r=1.0$, 1.56 and 100. Irrespective of $T$ and $h^{}_d$, $\mu^{}_oH^{}_c$ and $M^{}_r$ is negligibly tiny for square sample ($A^{}_r=1.0$). Even in the case of $A^{}_r=1.56$, there is a weak dependence of $\mu^{}_oH^{}_c$ and $M^{}_r$ on $T$ and $h^{}_d$ except at small temperature and dipolar interaction strength. Interestingly, there is an increase in $\mu^{}_oH^{}_c$ and $M^{}_r$ with $h^{}_d$ for a fixed temperature. There is a rapid fall in these values with temperature for weak dipolar interaction strength. While for sizeable dipolar interaction, there is a weak dependence of coercive field and remanence on temperature.} 
	\label{figure10}
\end{figure}

Finally, we study the variation of coercive field and remanence as a function of dipolar interaction strength and temperature for various values of aspect ratio. In Fig.~(\ref{figure10}), we plot the variation of $\mu_oH^{}_c$ (scaled by $H^{}_K$) and $M^{}_r$ as a function of $T$ and $h^{}_d$ for three representative values of aspect ratio $A^{}_r=1.0$, 1.56 and 100. The coercive field and remanence magnetization are negligibly small irrespective of $h^{}_d$ and $T$ for square sample. In the case of weak dipolar interaction and a large value of temperature, the coercive field and remanence have minimal values, which indicates the dominance of superparamagnetic character. The results with square sample is in qualitative agreement with the work of Figueiredo {\it et al.}~\cite{figueiredo2007}. We could not compare for other aspect ratio as they have concentrated on square geometry. It is also in qualitaive agreement with the work of Ewerlin {\it et al.}~\cite{ewerlin2013}. While for large dipolar interaction strength, there is a negligible value of remanence and coercive field as the antiferromagnetic coupling is dominant in this case. Similar observations can be made for $A^{}_r=1.56$ except at smaller dipolar interaction strength and temperature values where remanence and coercive field is  higher. Remarkably, the coercive field and remanence increase with an increase in $h^{}_d$ for a given temperature, provided the aspect ratio is huge. It is bacause of the enhanced ferromagnetic coupling. In the case of extremely large value of dipolar interaction strength, coercive field and remanence magnetization are very large and they depend very weakly on the temperature. The variation of $E^{}_H$ with $T$ and $h^{}_d$ is also found to be the same as that of coercive field variation. Therefore  we have not shown the corresponding curves of $E^{}_H$ variation.

\section{Summary and conclusion}
Now we summarize and discuss the main results of the present work. We have used kinetic Monte Carlo simulation technique to study the magnetic hysteresis in a two-dimensional arrangement of MNPs. We have systematically probed the hysteresis response as a function of dipolar interaction strength $h^{}_d$, system size ($L^{}_x\times L^{}_y$) or the aspect ratio $A^{}_r$, temperature $T$ and direction of the applied magnetic field. In the absence of thermal fluctuations and negligible dipolar interaction, the hysteresis curve obeys the Stoner and Wohlfarth model irrespective of aspect ratio and direction of the applied magnetic field. The superparamagnetic behaviour is dominant for weak dipolar interaction strength and large value of temperature, manifested in the minimal hysteresis loop area. Interestingly, the dipolar interaction is found to induce antiferromagnetic coupling provided the aspect ratio $A^{}_r$ is relatively small ($\leq6.0$) and $h^{}_d\geq0.2$, irrespective of the direction of the applied magnetic field. Therefore, the corresponding hysteresis curve shows a negligible value of remanence and coercive field~\cite{ewerlin2013}. On the other hand, for large dipolar interaction strength and $A^{}_r$, the ferromagnetic coupling is dominant with the magnetic field applied the sample's longer axis ($y$-direction). Temperature is also found to affect the hysteresis properties of the underlying system. In the case of weak dipolar interaction, the system moves from the blocked state to a superparamagnetic state as the temperature is increased. On the other hand, the temperature does not affect the hysteresis mechanism for large dipolar interaction strength.

The coercive field $\mu^{}_oH^{}_c$, remanence $M^{}_r$ and hysteresis loop area $E^{}_H$ are also useful quantifier to analyse the hysteresis properties of dipolar interacting MNPs. In the case of negligible and weak dipolar interaction, $\mu^{}_oH^{}_c$ and $M^{}_r$ are extremely small with a large value of temperature, irrespective of $A^{}_r$ and direction of the applied magnetic field. We can interprete it due to the dominance of superparamagnetic behaviour. Remarkably, the dipolar interaction is found to induce antiferromagnetic coupling between the magnetic moments for $h^{}_d>0.2$ and small values of $A^{}_r$ ($\leq6.0$). As a consequence, the corresponding hysteresis curves have negligibly small values of $M^{}_r$ and $\mu^{}_oH^{}_c$. When the magnetic field is applied along the long axis of the sample, there is an increase in coercive field and remanence with an increase in dipolar interaction strength for large $A^{}_r$, indicating the ferromagnetic coupling dominance. In the case of weak dipolar interaction, $\mu^{}_oH^{}_c$ and $M^{}_r$ decrease with an increase in temperature because of enhancement in thermal fluctuations. While for large dipolar interaction strength, they depend weakly on temperature. Remarkably, There is an intimate relationship between $E^{}_H$ and $\mu^{}_oH^{}_c$ as the variation of $E^{}_H$ with $h^{}_d$, $T$ and $A^{}_r$ is found to be the same as that coercive field variation.

In conclusion, we analysed the hysteresis properties in a two-dimensional assembly of dipolar interacting MNPs using kinetic Monte Carlo simulations. Our results suggest that the dipolar interaction affects the hysteresis mechanism drastically in such an array of MNPs. Interestingly, the nature of interaction can be tuned by varying the aspect ratio and direction of the applied magnetic field. We believe that these results are extremely useful for spintronics based applications. The observation made in the present article should help the experimentalists to manipulate the hysteresis response and other useful properties of interacting MNPs in a more controlled manner. These results should also be taken into account in interpreting of the experiments and efficient usage of magnetic
hyperthermia. Therefore, we hope that the present work could pave the way for joint efforts in experimental, theoretical, and computational
studies for these extremely useful and versatile systems.
%In the case of extremely large value of dipolar interaction stregnth coercive field and remanence magnetization are very large and they depend very weakly on the temperature.

%\section*{ACKNOWLEDGMENTS}

%\section*{DATA AVAILABILITY}
%The data that support the findings of this study are available from the corresponding author upon reasonable request.

\bibliography{ref}

\newpage

\newpage

\newpage

\newpage
\newpage

\newpage

\newpage
\newpage
\newpage
\newpage

\end{document}